# Isostatic Modelling, Vertical Motion Rate Variation and Potential Detection of Past-Landslide in the Volcanic Island of Tahiti

**Julien Gargani[1,2]**

[1]Univ. Paris-Saclay, CNRS, Geops, Orsay, France

[2]Univ. Paris-Saclay, Centre d'Alembert, Orsay, France

Corresponding author: Julien Gargani (julien.gargani@universite-paris-saclay.fr)

**Highlights**

- Large landslides could influence the vertical motion of the coastline when isostatic adjustment occurs
- Supposed stable volcanic islands used to calibrate sea-level change could be affected by isostatic rebound triggered by landslides.
- In Tahiti, a coastline uplift of 80-110 m occurred 872 kyr ago after a giant landslide
- A potential landslide that occurred 6 kyr ago could explain a subsidence rate variation of 0.1 mm/yr in Tahiti that affect sea-level curve reconstruction

**Abstract**

Intraplate volcanic islands are often considered as stable relief with constant vertical motion and used for relative sea-level curves reconstruction. This study shows that large landslides cause non-negligible isostatic adjustment. The vertical motion that occurred after landslide is quantified using a modelling approach. We show that a giant landslide caused a coastline uplift of 80-110 m for an elastic thickness of 15 km $< T_e <$ 20 km in Tahiti. Theoretical cases also reveal that a coastal motion of 1 m occurred for a landslide involving a displaced volume of 0.2 km$^3$ and influence relative sea-level reconstruction. In Tahiti, a change in the subsidence rate of 0.1 mm/yr (from 0.25 mm/yr to 0.15 mm/yr) occurred during the last 6 kyr and could be explained by a landslide involving a minimum volume of 0.2 km$^3$, 6 ± 1 kyr ago.

# 1 Introduction

Considerable research efforts have been dedicated to understanding the potentially destructive power of landslides, one of the principal processes responsible for relief destruction. Numerous landslides occur in a volcanic context (Clouard and Bonneville, 2004; Moore et al., 1995). The origin of landslides that occur on volcanic slopes could be related to volcanic processes (Cervelli et al., 2002; Clouard and Bonneville, 2004; Le Corvec and Walter, 2009). Nevertheless, other causes that can play a role in landslides on volcanic islands include sea level variations (Quidelleur et al., 2008). Although the causes of landslides have been extensively investigated (Carracedo et al., 1999; Crozier, 2010; Gargani et al., 2014; Hampel and Hetzel, 2008; Kilburn and Petley, 2003; Mc Mutry et al., 2001; Veveakis et al., 2007), their consequences on vertical movement are less studied.

The volume of rocks involved in a landslide event can vary by various orders of magnitude (Brunetti et al., 2009; Staron and Lajeunesse, 2007), some of which involve volumes equivalent to hundreds of cubic kilometres (De Blasio, 2020; Moore et al., 1995). Smith and Wessel (2000) suggested that landslides could cause isostatic rebound on the volcanic island of Hawaii. This phenomenon may have generated shoreline motion in Hawaii (Keating and Helsley, 2002) and in Canary Islands (Menendez et al., 2008) associated with other processes. Several large landslides have occurred in the Society archipelago (Clouard and Bonneville, 2004), and at least two large landslides involving volumes of 300 km$^3$ and 450 km$^3$ have occurred on Tahiti (Clouard et al., 2001; Clouard and Bonneville, 2004; Hildenbrand et al., 2006). Isostatic rebound after a landslide could be superimposed with other processes, such as rifting, thermal cooling, active volcanism (magma reservoir inflation/deflation), lithospheric flexure produced by the construction of a younger volcano, isostatic rebound after erosion, seismic displacement after a large earthquake, sedimentation and sea-level change (Ramalho et al., 2003).

Nevertheless, intraplate volcanic islands are considered stable relief submitted to constant vertical movement caused by volcanic edifice loading during several kyrs to million yrs (Lambeck, 2011) and used to estimate sea-level variations during the last 13 kyrs using coral reef (Pirazzoli and Montaggioni, 1988; Bard et al., 1996 and 2010; Hallmann et al., 2018). Absolute sea-level variation during the last million years are due to climatic variations caused by orbital forcing (Miller et al., 2005; Laskar et al., 2004) and more recently by anthropogenic contributions (Sahagian et al., 1994). Discrepancies between local sea level curves could be due to methodological problems, such as the underestimation of uncertainty, glacio-hydro-isostatic adjustments (Bard et al., 2016; Abdul et al., 2016), or variation in uplift or subsidence rates. Coral reef, that are located near sea surface, are used to estimate the position of the sea-level during the last kyrs through the datation and the measurements of accurate altitude or bathymetry of coral reefs. After the reconstruction of relative sea-level variations that depends from the local context, absolute sea-level variations curves are obtained by subtracting local vertical movements from the relative sea-level.

However, subsidence rates considered for the volcanic island of Tahiti, where a sea-level curve has been reconstructed, are not constant and are estimated at 0.2-0.25 mm/yr from 13.8 kyr to 7 kyr (e.g. Bard et al., 1996), at 0.15 mm/yr since 5 kyr (Hallmann et al., 2018; Pirazzoli and Montaggioni, 1988) and at 0.5 mm/yr at the present time using GPS measurements (Fadil et al., 2011), but no explanation is proposed to these differences. This study addresses the quantification of vertical motion caused by isostatic rebound following a major landslide in Tahiti and propose to interpret variation in vertical movement rates.

## 2 Geological and geomorphological setting

The island of Tahiti-Nui is a volcano formed between 1.4 Ma and 230 kyr in the Society archipelago (Clouard and Bonneville, 2005; Le Roy, 1994). The Society Island chain extends over 750 km from the present hotspot location under Mehetia, more than 100 km southeast of Tahiti, to the northwest (Duncan and McDougall, 1976; Munschy et al., 1998). The crust beneath the centre of the volcanic edifice of Tahiti is approximately 15-20 km thick (Calmant and Cazenave, 1986; Patriat et al., 2002) and decreases radially to a minimum thickness of 7 km away from the centre (Patriat et al., 2002). Lithospheric loading by the volcanic edifice weight is responsible for the subsidence of the island (Grevemeyer et al., 2001; Lambeck, 1981) and is estimated by reef age (radiocarbon and U-Th) and erosional features to have ranged from 0.15–0.39 mm/yr during the last 500 kyr (Pirazzoli et al., 1985; Thomas et al., 2012).

In the case of the northern Tahiti landslide that occurred 872 kyr ago (Fig.1), the length of the landslide deposits ranges from 50–80 km, and the maximum width of the landslide deposit is approximately 80–90 km, with a thickness of less than 500 m (Hildenbrand et al., 2006 and 2008). The landslide scar has a smaller width and length (20–25 km) than the landslide deposit that is located 30 km away and the initial volume before that any slide occurs has greater thickness (~2 km, Fig. 1B) than the landslide deposit. The significant spreading of the sliding material was caused by the dynamic of the Tahiti landslide that occurred catastrophically reaching high speeds >125 m/s (Gargani, 2020). Due to high runout and spreading, the landslide deposits are relatively thin and highly porous, whereas the destabilised material is composed of thick and dense volcanic material (Hildenbrand et al., 2006).

After the landslide, a subsequent volcanic eruption occurred and modified the morphology of the original scar (Le Roy, 1994). A new shield grew into the northern depression and overtopped the original volcanic structure around 500 kyr ago (Hildenbrand et al., 2006). When the infilling of the scar was full, the previous unloading by the landslide material removed from the scar was cancelled around 500 kyr ago. The present mean slope of the volcanic edifice ranges from 6–12°, as before the giant collapse (Hildenbrand et al., 2006; Ye et al., 2013).

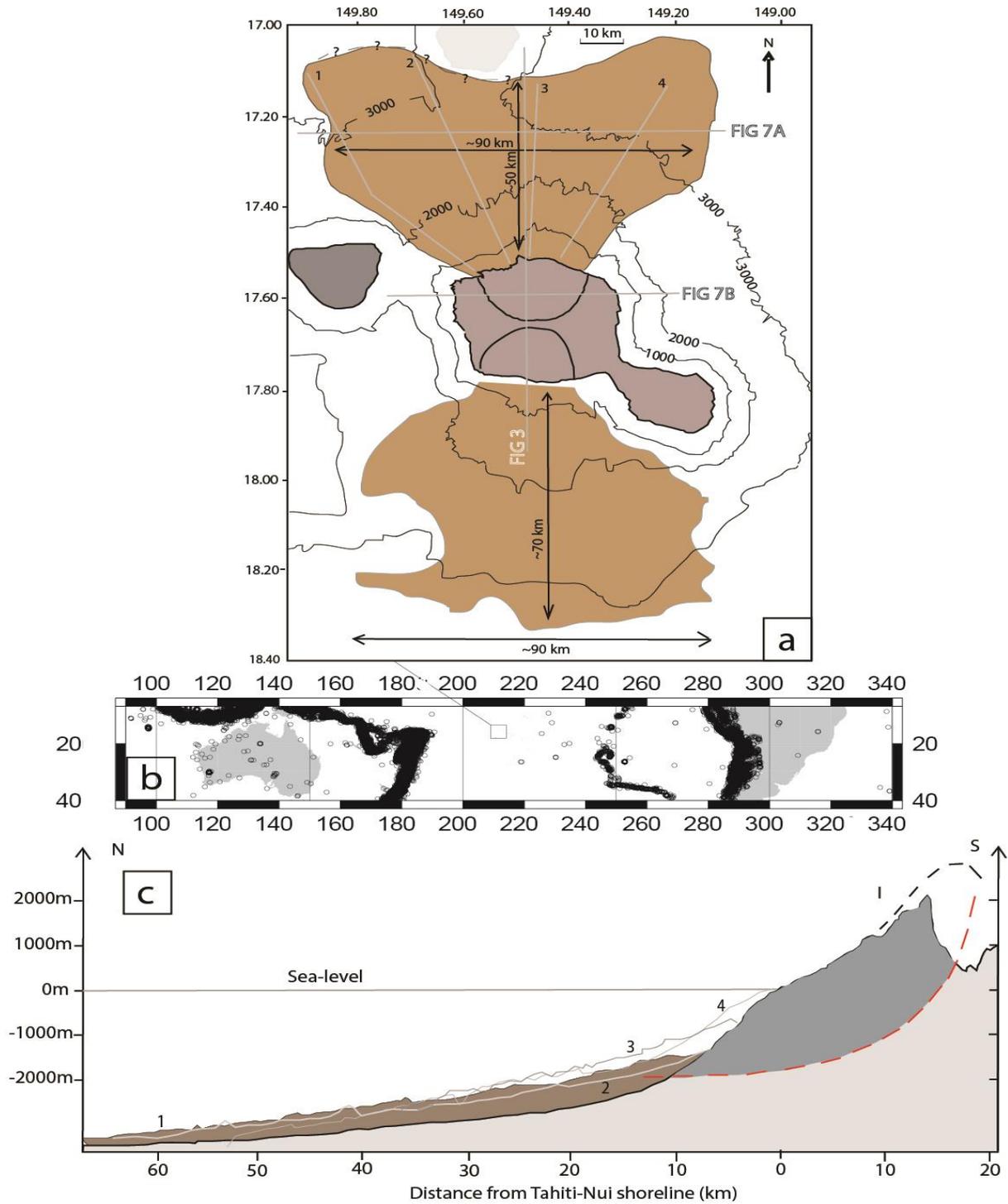

*Figure 1*. Map and section of Tahiti Island, French Polynesia: (**a**) The northern and southern landslide deposits on Tahiti Island are represented in brown. (**b**) Location of Tahiti in the Pacific Ocean and seismicity M>4. (**c**) The section of the northern landslide deposits is represented in brown for profile 1, and the infilling of the scar by volcanic material is in dark grey. The rotational

*landslide is in red. The locations of profiles 1–4 are shown in (**a**). Detailed data are provided in Clouard and Bonneville (2004) and Hildenbrand et al. (2006).*

## 3 Methods

A modelling approach is used to simulate the vertical movement of the lithosphere caused by a landslide. This study focus on the quantification of the effect of the isostatic rebound due to the landslide. Potential long-term uplift or subsidence caused by other processes could be superimposed with the isostatic rebound due to the landslides, but are not simulated. Concerning the vertical movement in an intraplate position, the initial conditions before any sliding could be considered in a steady state. The landslide transformed the lithosphere in isostatic equilibrium to a configuration that is not in isostatic equilibrium. Consequently, in elastic modelling isostatic restoring stresses immediately acted to regain equilibrium (Watts, 2001). Isostatic adjustment has been calculated using a classical 2D elastic model of the lithosphere (Gargani, 2004a, 2010; Weissel and Karner, 1989). The lithosphere is considered at isostatic equilibrium before loading or unloading caused by the landslide because no new loading by a volcanic edifice or eruption occurred in Tahiti during the last 200 kyrs. No horizontal forces are considered, and there is no vertical movement at the boundary of the model. The model is sufficiently large to let the area of interest be independent of the boundary condition. In the classical model of a thin infinite elastic plate, flexural isostasy $w(x,y)$ is implemented in a finite element code, with a 10 m step, using the equation (Turcott and Schubert, 2001):

$$\nabla^2(D \cdot \nabla^2 w(x,y)) + (\rho_a - \rho_v) g \cdot w(x,y) = \rho_v g [z_{ref}(x,y) - z(x,y)]$$

where $z_{ref} - z$ is the missing material thickness after the landslide, $\rho_a$ and $\rho_v$ are the densities of asthenosphere and volcanic rock respectively, $g$ the acceleration of gravity and $D$ is the rigidity that is calculated considering that :

$$D = E T_e^3 / [12(1-\nu^2)]$$

where $E$ is Young's modulus, $T_e$ is the effective elastic thickness, and $\nu$ is Poisson's ratio. The flexural isostasy equation is solved using a finite differences method. Poisson's ratio is considered constant at 0.25. In the modelling, the volcanic material and the crust have a density of 2800 kg/m³, whereas the density of the mantle is 3300 kg/m³ in agreement with estimates by Clouard et al. (2000). The density of water is 1000 kg/m³. The rigidity $D$ of the lithosphere generally ranges from $10^{21}$–$10^{23}$ Nm. On the volcanic islands of Tahiti, the elastic thickness estimated using the shape of the crust-mantle boundary ranges from $T_e = 15$ km (Patriat et al., 2002) to $T_e = 20$ km (Calmant and Cazenave, 1986), which corresponds to a lithospheric rigidity ranging from $D = 10^{22}$–$10^{23}$ Nm, respectively. For Tahiti, Lambeck (1981), Cazenave et al. (1980), Goodwillie and Watts (1993), and Grevemeyer et al. (2001) provided elastic thickness ($T_e$) values of 15, 20.9, 22.5, and 25 km, respectively. In this study, elastic thicknesses of 10, 15 and 20 km have been simulated.

In real world, the vertical movements associated with isostatic rebound take several thousand years to completely relax due to the viscous properties of the lithosphere. A duration of 5–15 kyr can be expected, as suggested by the duration of the isostatic rebound observed in Scandinavia and North America after ice sheet melt (10-15 kyr; van der Wal et al., 2010) and isostatic adjustments associated with deglaciation (5–7 kyr; Mitrovica et al., 2000). Here, a duration of 10 kyr to restore equilibrium through a vertical motion is considered. Consequently, vertical motion rates are estimated considering a linear vertical displacement during 10 kyr after the loading or unloading.

The first set of simulations considered unloading by the removed material and loading by landslide deposits of Tahiti giant landslide that occurred 872 kyr ago. In a 2D model implemented to calculate vertical motion resulting from isostatic adjustment, the geometry considered is a cross-sectional surface rather than a 3D volume (Fig. 2). The unloaded triangular landslide surface is calculated using the equation S=BxH/2, where B is the base of the triangular surface, and H is the height (Fig. 2A).

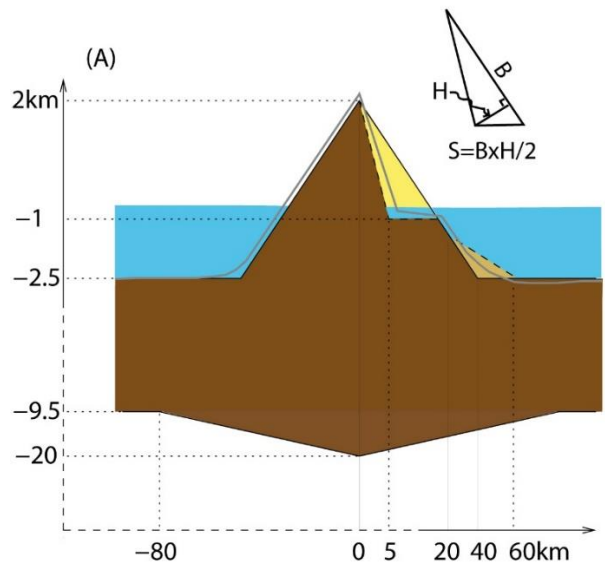

*Figure 2. Geometry of the volcanic island and of the unloading material in yellow considered for modelling the isostatic rebound. (A) The destabilised and deposited surfaces are illustrated in yellow and light brown, respectively. The crust is up to 20 km thick. The grey line represents the topography after isostatic rebound. Not to scale. (B) Various landslide geometries were simulated and are represented by dotted/dashed lines. The grey line represents the topography of the volcano after the landslide taking into account the isostatic rebound. (C) Influence of the sea level lowering on isostatic rebound. Not to scale. In (A), the horizontal scale is not homogeneous to allow to draw the base of the crust and the volcano on the same figure and consequently the landslide deposits (brown triangle) seem smaller than the landslide scar (yellow triangle).*

A North-South section was constructed considering the geometry of the volcanic edifice and the depth of the crust at Tahiti. To obtain realistic results, the geometry of a landslide that occurred approximately 872 kyr ago in the north of Tahiti Island was used. A volcano with a height of 4.5 km, width of 80 km, and mean slope of 6° was considered before the landslide that has been increased by erosional and landslide processes. Part of the volcano is located below the sea level, with only the highest 2 km of the topography of the volcanic edifice above sea level (Fig. 2A). The crust is approximately 20 km thick beneath the centre of the volcanic edifice and 7 km thick (from a depth of -2.5 km to a depth of -9.5 km; Fig. 2A) at 80 km from the centre of the volcanic edifice. The material removed during the landslide has a triangular geometry (Fig. 2A). The surface of the destabilised material could reach until $50 \times 10^6$ $m^2$ and was removed from the volcanic edifice to quantify the influence of the unloading on deformation (Fig. 2A and B). The removed material is composed of volcanic rock (Hildenbrand et al., 2006) and is described in the model using volcanic rock property for the density.

In all the simulations, the shoreline position corresponds to the position of the altitude z = 0 m along the simplified topography of the volcano. The uplift presented is the one at the position of the shoreline before the landslide. The destabilised material was modelled between 0–20 km laterally from the summit (Fig. 2A). The loading of the volcanic material accumulated at the base of the volcanic edifice due to the landslide deposits spans from x = 20–60 km and has a triangular geometry (Fig. 2A). The cross sections of the destabilised and deposited material are triangular and approximatively correspond to the landslide that occurred 872 kyr ago in Tahiti. The thickness of the unloaded material is higher than the thickness of the material deposited by the landslide. The spreading of the landslide deposits generated a relatively thin deposit across a wide area (Fig. 1).

In a second set of simulations, isostatic response was implemented to calculate the effect of unloading with different mass and geometry for the landslide while neglecting landslide deposit loading. Various triangular geometries for the landslide section from those comparable to large landslides (>20x10$^6$ m$^2$) to those of more usual ones (0.2x10$^6$ m$^2$) were tested (Fig. 2B, Table 1). For the landslide ranging from 2.5x10$^6$–5x10$^6$ m$^2$, the post-landslide slope was ~9° in the landslide scar area. The deposits geometry is difficult to evaluate because variable spreading and runout could take place. The precise spreading depends on the geological context (water or not, lithology, caused by an explosive volcanic eruption or not…), but also on the dynamic of the landslide itself (slow vs catastrophic). A realistic "mean" geometry is difficult to quantify. Nevertheless, landslide mobility is generally high, and spreading is significant for large landslides (Staron and Lajeunesse, 2009; De Blasio, 2020), reducing the effective weight of the deposit column in comparison to the thick initial material displaced from the destabilized area. In Tahiti, the material could travel on high distances (Gargani, 2020) and this is why the loading is neglected. When the mobility and the spreading of the material became infinite, the loading by deposited material became negligible. When the loading of the deposits is not considered, the uplift is overestimated by a small amount in the event of significant mobility and spreading of the landslide.

| Landslide section (m$^2$) | Elastic thickness (km) |
|---|---|
| 0.2x10$^6$ | 15 |
| | 20 |
| 0.45x10$^6$ | 15 |
| | 20 |
| 1x10$^6$ | 15 |
| | 20 |
| 4x10$^6$ | 15 |
| | 20 |
| 7x10$^6$ | 15 |
| | 20 |
| 20x10$^6$ | 15 |
| | 20 |
| 45x10$^6$ | 15 |
| | 20 |

*Table 1: Second set of experiments.*

## 4 Results

The first set of simulations allowed us to evaluate the interaction between the area where unloading and uplift occurred with the area where loading by the deposited material and subsidence occurred. Modelling results suggest that isostatic deformation affected larger areas than the areas where loading and unloading occurred, due to the mechanical flexure of the lithosphere. It can be observe that the modelled uplift caused by unloading is higher than the modelled subsidence caused by loading (Fig. 3a and b). The extent of the area affected by uplift is larger than the area affected by unloading, because the spreading of the deposited material reduces the column weight and also because flexural rigidity distribute the deformations also laterally apart from the unloaded/loaded column. A length of at least 70 km was affected by uplift due to the flexural rigidity of the lithosphere. In the modelling, the amplitude of the isostatic adjustment principally depends on the mass displaced (i.e. the section of surface removed or superimposed) and on the flexural rigidity $D$ (i.e. the elastic thickness $T_e$) of the lithosphere. Lower rigidity allows more uplift (Fig. 3 and 4). Modelled uplift occurred on the volcano flanks above and below sea level (Fig. 3). At an altitude of 0 m—the supposed initial coastline level—there are uplifts of 80 m ($D = 10^{23}$ Nm, i.e. $T_e = 20$ km), 110 m ($D = 10^{22}$ Nm, i.e. $T_e = 15$ km), and 190 m ($D = 10^{21}$ Nm, i.e. $T_e = 10$ km) (Fig. 4a). The maximum uplift ranges from 85 m to 200 m and its position depends on the elastic thickness (Fig. 3b).

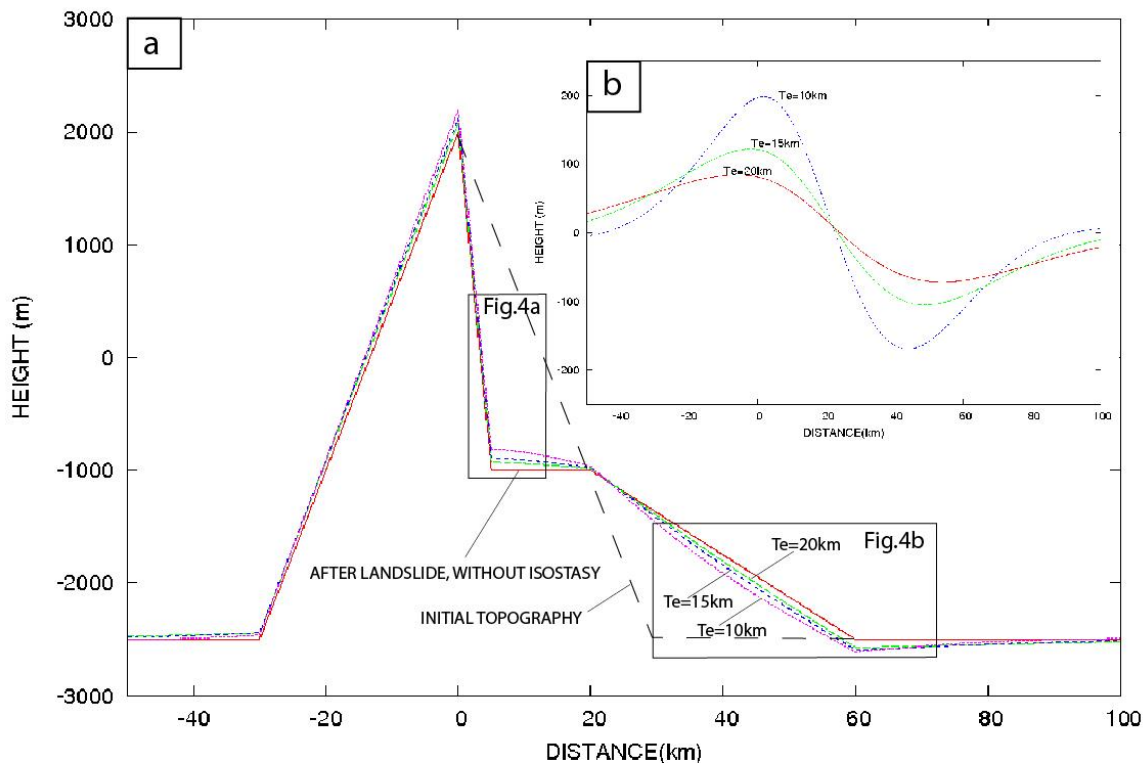

*Figure 3*. Vertical movements and unloading effects. (***a***) Isostatic adjustment after the large Tahiti landslide for three different elastic thickness. (***b***) Absolute vertical movement due to isostatic uplift and subsidence on Tahiti after the large landslide. Three different elastic thickness ($T_e$) values

*were used (10 km; 15 km; 20 km) to simulate the vertical movement caused by isostatic response after the large landslide.*

Due to loading by landslide deposits, a maximum subsidence of 70 m ($T_e = 20$ km), 100 m ($T_e = 15$ km) and 160 m ($T_e = 10$ km) occurred at the foot of the volcano (Fig. 3b). The extent of the subsidence depends on the flexural rigidity.

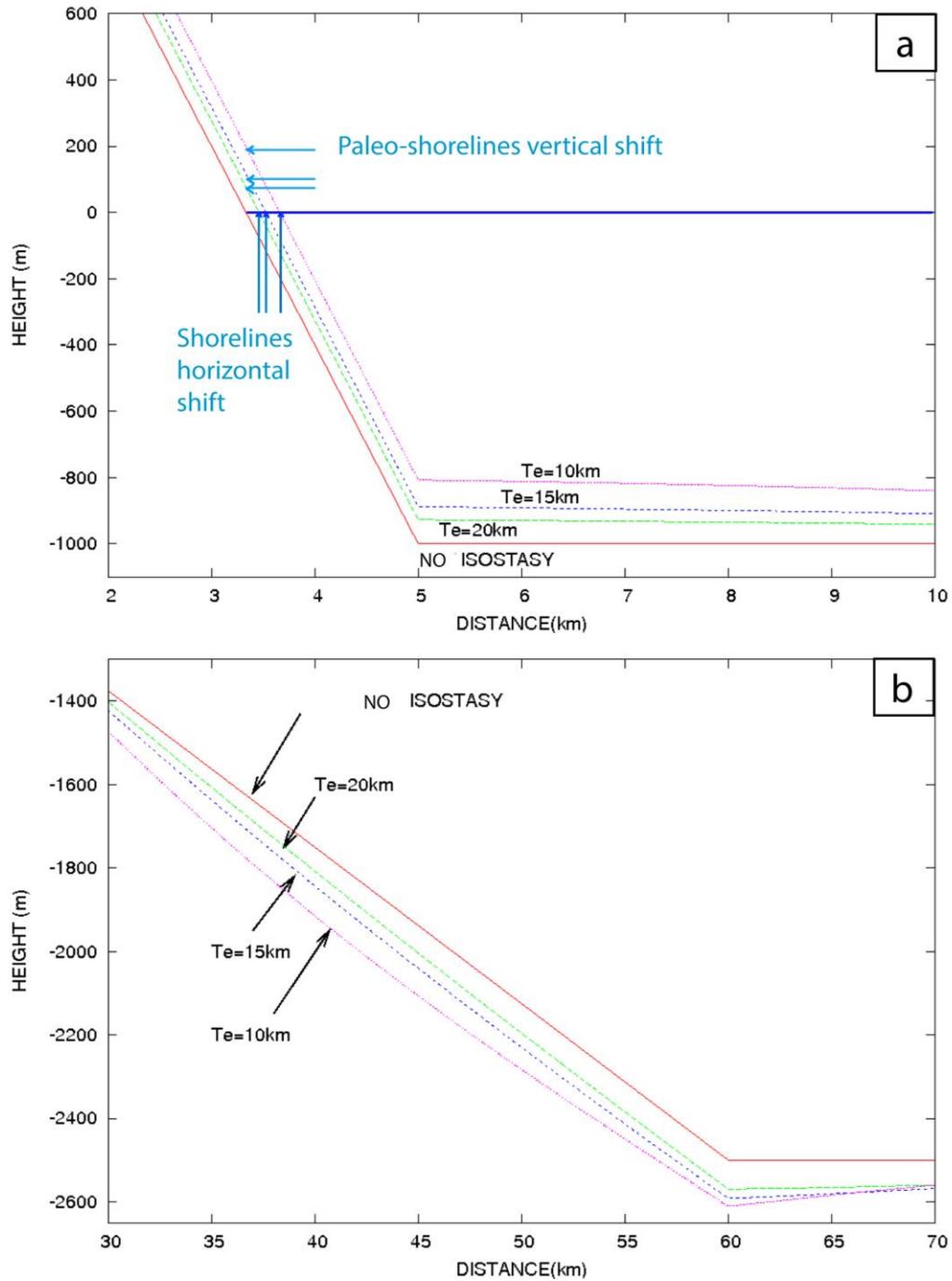

*Figure 4. Insert of figure 3. Positions of the surface of the volcano after isostatic adjustment with*

*different $T_e$. Details of vertical movements for $T_e = 10$ km, $T_e = 15$ km, and $T_e = 20$ km. (**a**) Isostatic adjustment of the coastline after the large landslide that occurred on Tahiti approximately 872 kyr ago. The relict shorelines after isostatic adjustment are represented by blue horizontal arrows for each elastic thickness $T_e$. The new shorelines, after that isostatic response take place, are represented by blue vertical arrows. (**b**) The detailed distal part of the volcano where the landslide deposits are located and subsidence occurred. The red line represents the topography of the volcanic island without isostatic deformation. The slope around the coast in the area where the landslide occurred is ~12°.*

Large landslides (>20x10⁶ m²) generate significant deformations (in the case of fig. 3 and 4, the displaced cross-sectional surface was of 22.5x10⁶ m²), and moderately sized landslides (2.5x10⁶ m² < S < 5x10⁶ m²) also cause non-negligible uplift (Fig 6a and b), which depends on the cross-sectional surface affected by the landslide and the rigidity of the crust. Coastal uplift ranges from 0 m to approximately 160 m for a landslide surface of 0–50x10⁶ m². The calculation has been done for various volume that are summarized in table 2. More precisely, for a large landslide with a cross-sectional surface removed of 50x10⁶ m² (i.e. a displaced volume of 1000 km³ assuming a width of 20 km), there is 160 m of uplift for elastic thickness $T_e = 15$ km and 135 m of uplift for $T_e = 20$ km (Table 2). For a cross-sectional surface removed of 10x10⁶ m² (i.e. ~50 km³ displaced volume considering a mean width of 5 km), the landslide cause a coastal uplift of 50 m. A landslide with a cross-sectional surface removed of 0.2x10⁶ m² (i.e. ~0.2 km³ displaced volume considering a mean width of 1 km) produces an uplift of 1 m. This last result is comparable with a sea-level lowering of ~4 m that also cause an isostatic rebound of 1 m (Fig. 5).

| Landslide surface (section in m²) | Landslide Volume (km³) | Landslide Width (km) | Elastic thickness $T_e$ (km) | Coastal Uplift (m) |
|---|---|---|---|---|
| 0.2x10⁶ | 0.2 | 1 | 15 | 1.1 |
| 0.2x10⁶ | 0.2 | 1 | 20 | 0.9 |
| 10x10⁶ | 50 | 5 | 15 | 50 |
| 10x10⁶ | 50 | 5 | 20 | 40 |
| 50x10⁶ | 1000 | 20 | 15 | 160 |
| 50x10⁶ | 1000 | 20 | 20 | 135 |

*Table 2: Summarizing main modelling results.*

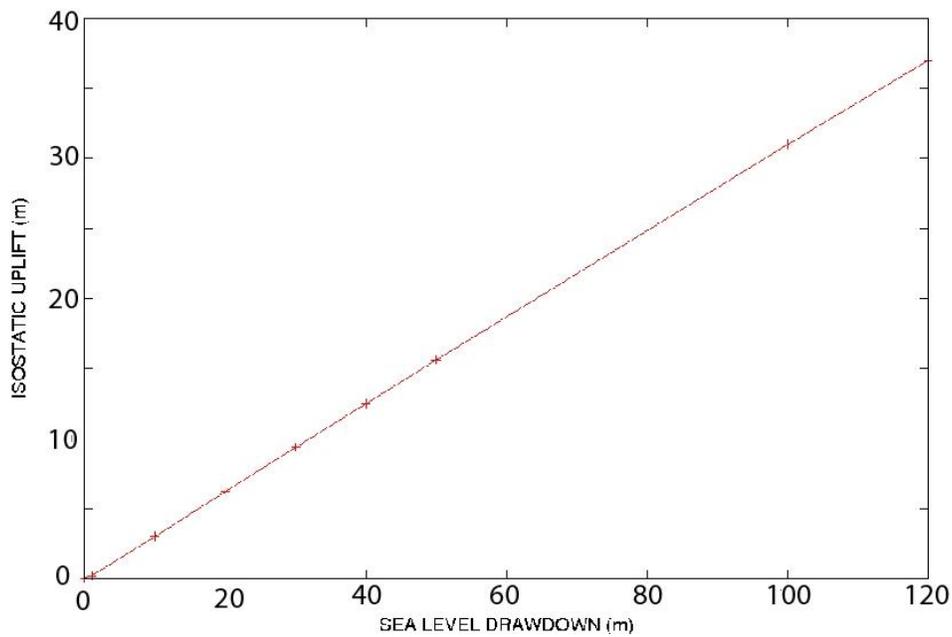

*Figure 5. Coastal uplift due to isostatic rebound triggered by sea level unloading. $T_e$ =15 km.*

A relatively "small" landslide (S = $0.2 \times 10^6$ m$^2$ equivalent to 0.2 km$^3$ displaced volume) that cause a 1 m uplift in approximately 10 kyr generates a mean uplift rate $V_U$ of 0.1 mm/yr (Fig. 6b). This is 100 times smaller than the uplift rate caused by a giant landslide as the one that occurred 872 kyr ago in Tahiti. In this case, the expected uplift is of ~100 m and the resulting mean uplift rate is of $V_U$ = 10 mm/yr (Fig. 6a).

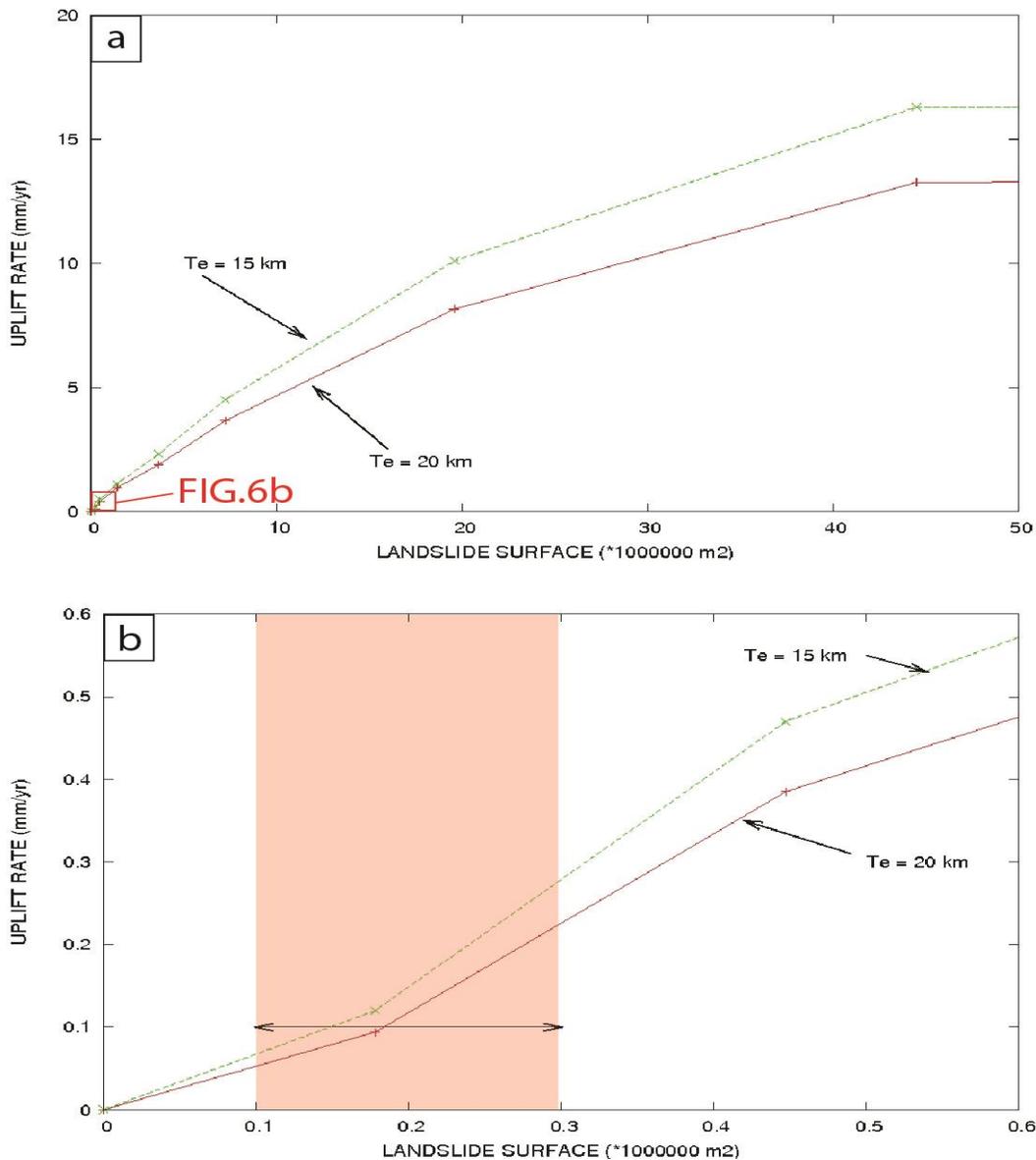

*Figure 6*. Vertical motion rate after a landslide due to isostatic adjustment. (**a**) Influence of the cross-sectional surface of the landslide on the uplift rate considering that the isostatic rebound take place in 10 kyr, (**b**) Detail of the uplift rate for landslide cross-sectional surfaces $< 0.6 \times 10^6$ $m^2$. An uplift rate of 0.1 mm/yr, equivalent to the variation of the subsidence rate in Tahiti, is highlighted by a double arrow. The landslides able to cause an uplift of 0.1 mm/yr ranges from $0.1 \times 10^6$ $m^2$ to $0.3 \times 10^6$ $m^2$

The isostatic adjustment is also calculated perpendicularly to the one presented in figure 3,4 and 6 and is presented in figure 7. More precisely, in figure 7B, the isostatic rebound is modelled in the area of the landslide scar (see figure 1A for location) and uplift is estimated. In figure 7B, the isostatic adjustment is modelled where the landslide deposit can be observed (see Figure 1A for location) and significant subsidence is estimated.

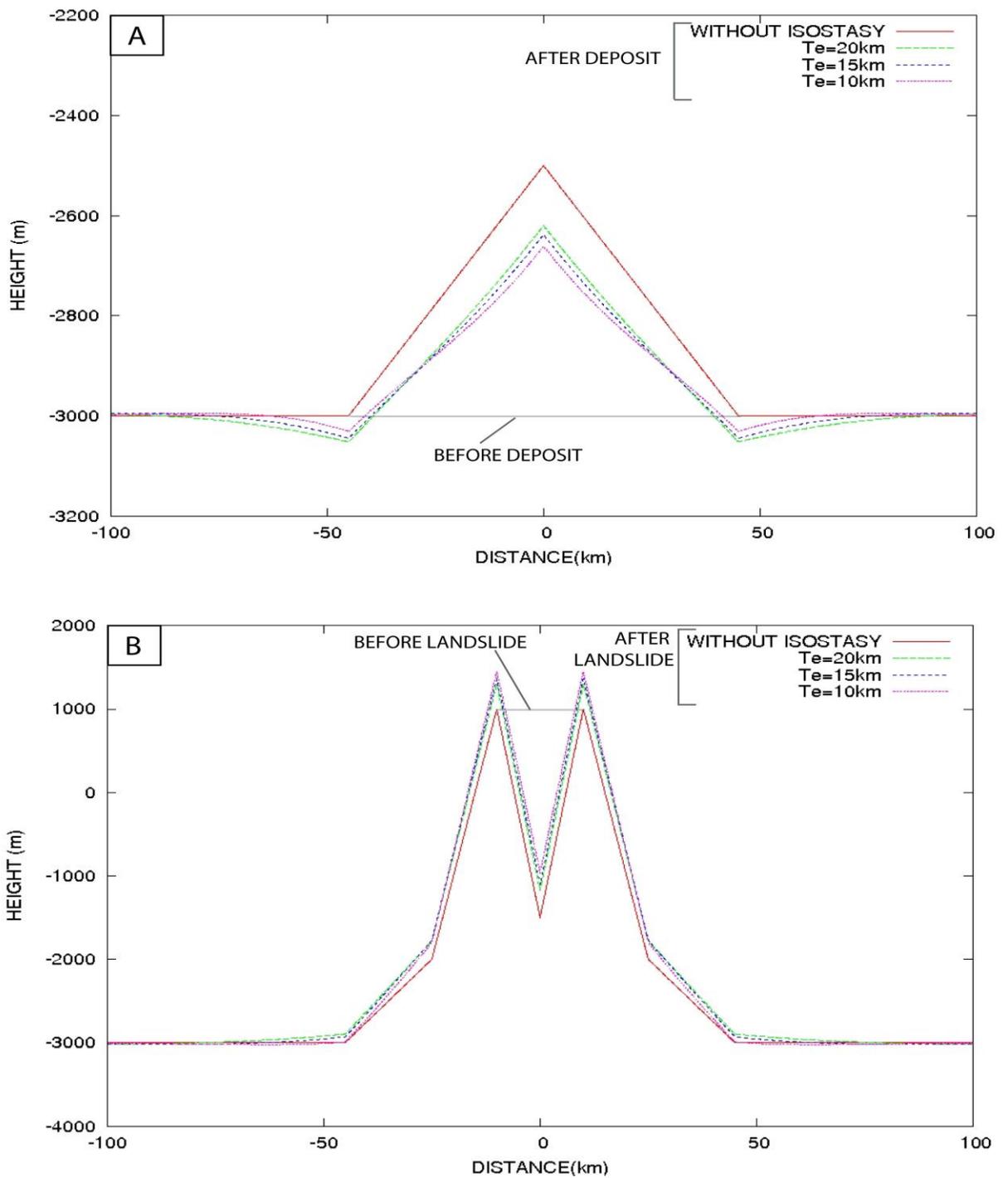

*Figure 7*: Modelling of the isostatic adjustment on the Tahiti volcano after a giant landslide (a) in the deposit area, (b) in the landslide scar area. These models are perpendicular to the previous ones presented in figures 3, 4 and 6. The position of these sections is showed in figure 1A.

# 5 Discussion

## 5.1 Effect of a giant landslide on vertical motion

The geomorphologic changes associated with isostatic rebound should have been significant in Tahiti after the 872 kyr giant landslide. Modelling results suggest that the shoreline uplifted between 80 and 110m for a $T_e = 20$ km and for a $T_e = 15$ km, respectively. This is comparable with results obtained in Hawaii for a 1200-5000 km$^3$ volume slides considering $T_e = 25$ km (Smith and Wessel, 2000). When the coast is uplifted, the relicts of marine terraces are visible (Binnie et al., 2016), or/and cliffs can form on steep coasts (Martinod et al., 2016). High littoral cliffs circling two-thirds of Tahiti with an elevation ranging from 60 m at the East, approximately 100 m in the North, and approximately 200 m in the West and South could be observed but are considered to have been edified during the last 120 kyrs (Ye et al., 2013) and can't be related to this giant landslide. Nevertheless, the loading by the formation of a second shield (Hildenbrand et al., 2006) have caused a subsidence that annihilate the previous uplift and the morphological evidence should be offshore.

## 5.2 Effect of a moderate landslide on vertical motion

At least 39 landslides have been described in the Society Islands and the Austral volcanic archipelago with various volume involved (Clouard and Bonneville, 2004). The significant number of landslide recorded in the Society Islands suggests that volcano flank destabilisation is a current phenomenon in this area. Small landslides recently occurred in Tahiti where high precipitation rates and weathered rock can be observed (Guillande et al., 1993; Parkes et al., 1992). Statistically, it have been showed that the number of landslides increases as their volumes decreases (Brunetti et al., 2009). Landslides involving a volume of ~0.2 km$^3$ are relatively current events and should have caused vertical motion rate variation of ~0.1 mm/yr.

An indirect observation of a landslide is possible using subsidence rate in Tahiti obtained with coral reef age and elevation or depth for sea-level curve reconstruction (Fig. 8). There are some discrepancies between the subsidence rate estimates during the last 5 kyr from the one estimates for 13.8-7 kyr. Pirazzoli et al. (1985) suggest a subsidence rate of 0.15 mm/yr during the last 5 kyr, whereas Bard et al. (1996) consider a constant value of 0.2-0.25 mm/yr from 13.8 kyr to 7 kyr to calibrate the local sea-level curve reconstructed using coral reef. These subsidence rates considered by Pirazzoli et al. (1985), Hallmann et al. (2018) and Bard et al. (1996) are necessary to compare Tahitian sea-level curves with other local sea-level curves and permit the coherency of the sea-level reconstruction all over the Earth (Abdul et al., 2016; Bard et al., 1996). The comparison of the sea-level curves from Pirazzoli et al. (1985), Hallmann et al. (2018) and Bard et al. (1996) suggests that a slowdown of 0.1 mm/yr occurred since $6 \pm 1$ kyr (Figure 8).

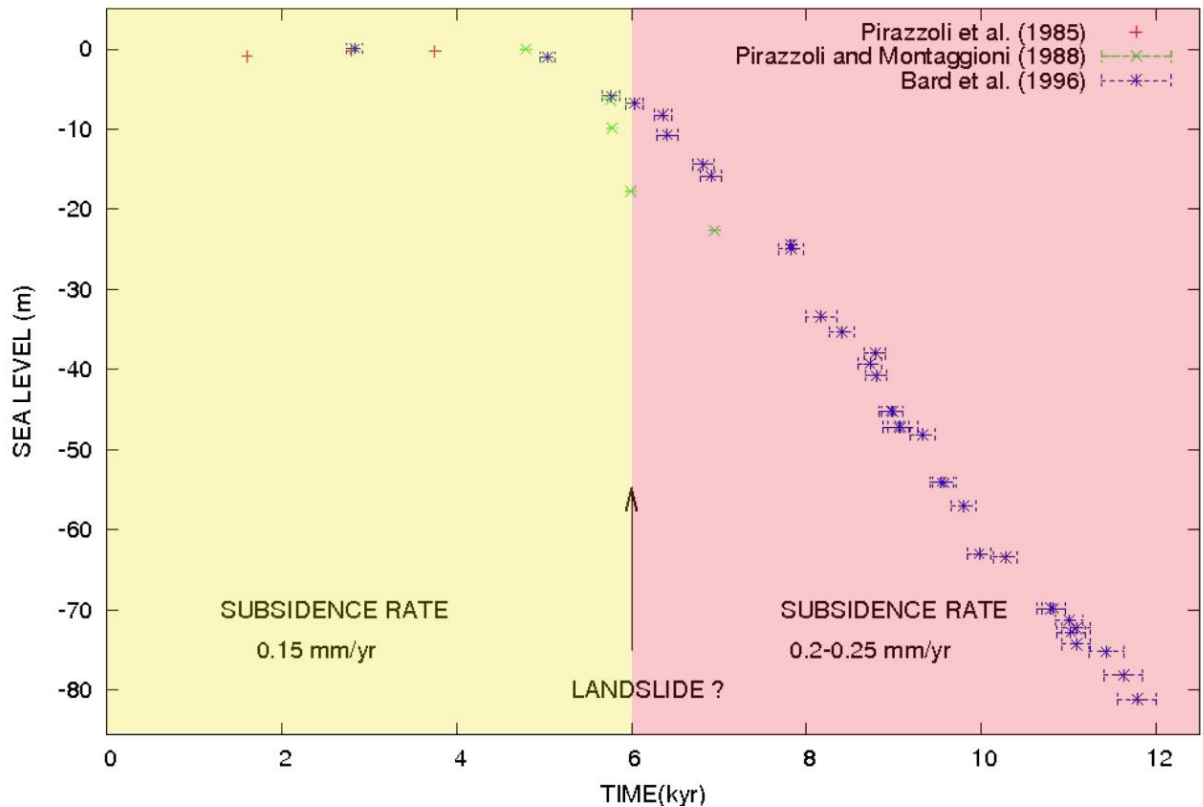

*Figure 8. Sea level and subsidence rates in Tahiti. Sea level variations were estimated by Pirazzoli et al. (1985), Pirazzoli and Montaggioni (1988) and Bard et al. (1996). For Bard et al. (1996) only samples dated by $^{14}C$ are represented for clarity.*

When a transient (~10 kyr) uplift rate $V_U$ (with $V_U > 0$ by convention) caused by a landslide is contemporaneous with a significant long-term subsidence context (subsidence rate $V_S < 0$ by convention), the observed vertical motion rate became $V_O = V_S + V_U$ and $V_O < 0$ when the subsidence rate is higher than the transient uplift rate caused by the isostatic adjustment. A landslide with a removed cross-sectional surface of $0.2 \times 10^6$ m$^2$ (i.e. $> 0.2$ km$^3$ for a width of 1 km) cause an uplift rate $V_U = 0.1$ mm/yr (Fig. 6b) that could explain the subsidence rate slowdown of 0.1 mm/yr that occurred since $6 \pm 1$ kyr at Tahiti. It suggest the occurrence of a landslide $6 \pm 1$ kyr ago with a volume of 0.2 km$^3$ at Tahiti. This is a minimum value for the landslide size, because the deposits were considered to spread infinitely. It would have been the most important and the unique event of this size in Tahiti.

The cliff observed at Tahiti have been explained by a relative sea-level fall of +5m, 7 kyrs ago (Ye et al., 2013), but any uplift in Tahiti has been recorded by coral reef from 13.8 to 1.6 kyr. Cliffs retreat could be interpreted by processes involving erosion, sea-level variation and vertical movements. However, it is difficult to separate the influence between these parameters. Low atmospheric pressure associated with hurricane cause storm surges and waves of several meters above sea-level (Rey et al., 2019). Storm surges and waves could generate marine erosion above sea-level. It could be difficult to discriminate these morphological features from those caused by eustatic variations.

*5.3 Sea-level variation and water loading hypothesis*

Other processes could influence vertical movement rates such as (*i*) lithospheric flexure caused by the loading of a distant new volcano or the load of a new eruption, (*ii*) isostatic rebound after a sea-level lowering or (*iii*) a massive erosion, tectonic uplift or subsidence. These potential effects are discussed hereafter.

A sea-level lowering of less than 1 m has been recorded during the last kyr at Tahiti (Hallmann et al., 2018; Pirazzoli and Montaggioni, 1988). Hallmann et al. (2018) suggested a sea level high stand of +0.9 m from 3.9–3.6 kyr before present in French Polynesia to interpret the marine relict located above present sea-level. A sea-level lowering of 1 m generates isostatic uplift of approximately 0.25 m (Fig. 5) —equivalent to an uplift rate of 0.025 mm/yr assuming that the equilibrium is reached in ~10 kyr. As a consequence, the unloading due to recent sea-level lowering caused a transient uplift rate $V_U = 0.025$ mm/yr and cannot explain the decreasing subsidence rates in Tahiti from 0.25 mm/yr between 13.8–7 kyr (Bard et al., 1996) to 0.15 mm/yr (Hallmann et al., 2018; Pirazzoli and Montaggioni, 1988) during the last ~6 kyr.

*5.4 Erosion and isostatic response*

Other phenomenon could also produce isostatic rebound and could be superimposed with subsidence rates caused by volcano loading. Several studies suggested that erosion could generate isostatic uplift (Gargani, 2004b and 2010; Menendez et al., 2008), for example in the Canary volcanic Island (Menendez et al., 2008). In Tahiti, regressive erosion is non-negligible (Ye et al., 2013). It is well known that regressive erosion is favoured by sea-level lowering, uplift or climatic change (Gargani, 2006 and 2010; Loget and Van den Driessche, 2009). Nevertheless, no abrupt climate change or a significant (> 5 m) sea-level lowering have been documented in Tahiti during the last 5-7 kyr (Bard et al., 1996; Hallmann et al., 2018; Pirazzoli and Montaggioni, 1988). Erosion in Tahiti take place since a long time, just after the formation of the new shield into the giant landslide scar around 500 kyr ago (Hildenbrand et al., 2006). If no major climatic or sea-level fall change take place in Tahiti since 6 kyr, no major erosion change could have taken place and explain the variation of the subsidence rate.

*5.5 Volcano loading and vertical motion*

A lithospheric loading by a new volcanic edifice could have modified the flexure and changed the subsidence rate. Around 100 km at the south-east of Tahiti, Mehetia Island is dated 70-75 kyr at the base of its stratigraphic piles (Binard et al., 1993), but younger events are recorded until 3 kyr (Binard et al., 1993; Clouard and Bonneville, 2005). 50 km at the east of Tahiti, Teahitia seamount age is estimated to be between 380 kyr and 50 kyr old (Clouard and Bonneville, 2005). However, Teahitia seamount is still active (German et al., 2020). It cannot be excluded that the growth of Mehetia and Teahitia volcanos could have influenced the flexure of the lithosphere beneath Tahiti during the Holocene. An elastic thickness of 10-20 km, compatible with a crust thickness of 12-20 km beneath volcanos in the Society archipelago (Calmant and Cazenave, 1986; Sichoix, 1997; Patriat et al., 2002), could generate flexure on distance > 100 km. Nevertheless, the amount of volcanic material potentially accumulated during the last 6kyr on Mehetia Island and Teahitia seamounts is still unknown and its influence is negligible comparatively to other processes nearer to Tahiti.

Other processes such as a permanent deformation associated with strong earthquakes are not appropriate to explain the change in the subsidence rate of Tahiti. Tahiti is located in an intraplate area where seismicity with a magnitude M>4 is not usual. Volcano-seismic activity is located more than 40 km away from Tahiti at Teahitia (German et al., 2020) and more than 100 km away from Tahiti at Mehetia (Talendier and Okal, 1984).

*5.6 Triggering mechanism of paleo-landslides*

The potential landslide that may have taken place 6 kyr ago should have occurred after a significant sea-level rise (Fig. 8). The consequences of a water-level increase on the triggering of landslides have been evidenced in the case of the Vajont landslide (1963, Italy) (Muller-Salzburg, 1987; Kilburn and Petley, 2003; Veveakis et al., 2007), suggested in the case of volcanic islands (McMutry et al., 2004; Quidelleur et al., 2008) or on continental margins (Gargani et al., 2014). Contemporaneously with Quaternary sea-level rise, climatic conditions use to change. Increase of precipitation causes pore pressure increase and triggers landslide (Cervelli et al., 2002). Highly weathered rocks by wet climatic conditions causes the weakening of geomechanical parameters (cohesion, angle of friction) and favor volcano flank destabilization (Rodriguez-Losada et al., 2009). Alternatively, volcanic activity such as pressure reservoir variation could have caused deformation and slip on preexisting fault (Gargani et al., 2006b) favoring slope destabilization.

## 5 Conclusions

Coastal uplift generated by isostatic uplift after a landslide was quantified. Landslides generate non-negligible isostatic adjustment on intraplate volcanic islands. Coastal uplift ranges from 1–50 m for landslide surfaces ranging from $0.2 \times 10^6$–$10 \times 10^6$ m$^2$. A large landslide of approximately $22.5 \times 10^6$ m$^2$, (i.e. 450 km$^3$ displaced volume), such the one that occurred on Tahiti Island approximately 872 kyr ago, generates a coastline uplift of 80–110 m depending of the flexural rigidity of the lithosphere. A potential landslide with a cross-sectional displaced surface of $0.2 \times 10^6$ m$^2$ (i.e. 0.2 km$^3$ displaced volume considering a landslide width of 1 km) could explain the variation of subsidence rate of 0.1 mm/yr observed since 6 ± 1 kyr ago at Tahiti. This method could permit to detect past-unknown landslides or to interpret unexplained variations in vertical motion rates.

**Acknowledgments:**
- Reviewers are acknowledged.

**Declarations of interest**:
- Author declare no conflict of interest with respect to the results of this paper.